\documentstyle[12pt]{article}
\begin{document}
\def\om{\omega}
\def\omt{\tilde{\omega}}
\def\ti{\tilde}
\def\o{\Omega}
\def\bchi{\bar\chi^i}
\def\In{{\rm Int}}
\def\ba{\bar a}
\def\w{\wedge}
\def\ep{\epsilon}
\def\k{\kappa}
\def\Tr{{\rm Tr}}
\def\ST{{\rm STr}}
\def\ss{\subset}
\def\ot{\otimes}
\def\bc{{\bf C}}
\def\br{{\bf R}}
\def\de{\delta}
\def\tr{\triangleleft}
\def\al{\alpha}
\def\la{\langle}
\def\ra{\rangle}
\def\G{\Gamma}
\def\th{\theta}
\def\lm{\lambda}

\def\jp{{1\over 2}}
\def\js{{1\over 4}}
\def\d{\partial}
\def\ds{\partial_\sigma}
\def\dt{\partial_\tau}
\def\d+{\partial_+}
\def\d-{\partial_-}
\def\be{\begin{equation}}
\def\ee{\end{equation}}
\def\bea{\begin{eqnarray}}
\def\eea{\end{eqnarray}}
\def\D{{\cal D}}
\def\G{{\cal G}}
\def\H{{\cal H}}
\def\E{{\cal E}}
\def\C{{\cal C}}
\def\bT{\bar{\cal T}}
\def\F{{\cal F}}
\def\n{{1\over n}}
\def\si{\sigma}

\def\e{\varepsilon}
\def\b{\beta}
\def\ga{\gamma}

\begin{titlepage}
\begin{flushright}
{}~
IML 99-21\\
hep-th/9906163
\end{flushright}

\vspace{3cm}
\begin{center}
{\Large \bf Supersymmetric gauged WZNW models as  dressing cosets}\\
[50pt]{\small
{\bf C. Klim\v{c}\'{\i}k and S. Parkhomenko}\footnote{Permanent address:
Landau Institute for Theoretical Physics, Chernogolovka, Russia}
\\ ~~\\Institute de math\'ematiques de Luminy,
 \\163, Avenue de Luminy, 13288 Marseille, France}

\vspace{1cm}
\begin{abstract}
The domain of applicability of the Poisson-Lie T-duality is enlarged
to include the gauged WZNW models.

\end{abstract}
\end{center}
\end{titlepage}
\newpage

1.T-duality in string theory is a subject which has been keeping
 attracting attention
in the last few years.  Apart from the well-understood Abelian T-duality
\cite{KY}, there persists a problem of interpreting the mirror symmetry
as a sort of T-duality \cite{SYZ}. In particular, the Calabi-Yau manifolds
do not have Abelian (or non-Abelian) groups of
isometries needed for applying the Abelian T-duality.
Obviously a framework is needed which would enable us to dualize the models
living on targets without necessarily requiring any isometry. A step
forward was made in the work \cite{KS1}, where the traditional non-Abelian
T-duality \cite{OQ} was generalized to the so-called Poisson-Lie
T-duality. The latter duality is based on a generalisation of the standard
isometries called Poisson-Lie symmetries and it basically replaces
one Poisson-Lie group (target space) with its dual Poisson-Lie group (dual
target). In a subsequent series of papers
\cite{KS2,KS3,KS4,K,P1,P2,Ost} the Poisson-Lie T-duality has been well
 developped in many of its
aspects; perhaps the most important contributions from the structural point
of view were two papers: \cite{KS3} and \cite{KS4}. In \cite{KS3} the
global aspects of the Poisson-Lie  T-duality have been addressed.It was
found that the duality takes place not only between the  pairs of the
 Poisson-Lie
groups but more generally between
 the pairs of coset spaces $D/G$ and $D/\ti G$
where $D$ is a so-called Drinfeld double and $G,\ti G$ is a pair of its
maximally
isotropic soubgroups. In \cite{KS4} the duality was further generalized
to incorporate pairs of double coset spaces $F\backslash D/G$ and
$F\backslash D/\ti G$, where $F$ is some isotropic subgroup of $D$ that
is not necessarily maximally isotropic. This double coset duality was
referred to as the dressing coset duality because in the case when
$D/\ti G=G$, $F$ indeed acts on $G$ in the dressing way.

An interesting aspect of the dressing cosets $F\backslash D/G$ is that they
 are not necessarily
nonsingular manifolds as the targets $D/G$. Moreover, there seems to be
no natural group action on the targets $F\backslash D/G$ as it is the case
for $D/G$. All this makes the dressing cosets to be sufficiently "wild"
objects in order to have some hope that they could include the Calabi-Yau
manifolds. However, at the present stage it is still a somewhat premature
discussion and in this note we wish to concentrate on another important
aspect of the Poisson-Lie T-duality: a conformal invariance of the
 corresponding
$\si$-models.

The issue of the conformal invariance has been always somewhat mysterious
in the framework of the Poisson-Lie T-duality which, so far, is mainly a
classical story. Nevertheless, it turned out that the WZW model on compact
groups, which is certainly conformal also at the quantum level, fits very
naturally in the framework of the single coset ($D/G\leftrightarrow
D/\ti G$) duality \cite{KS2}.
It is the purpose of this note to show that the gauged WZW models also fit
very well into the framework of the Poisson-Lie T-duality but at the
double coset level $F\backslash D/G$. We shall then coclude that the
Poisson-Lie T-duality concerns very directly the conformal backgrounds
and presumably also realistic string models.

We shall actually work out also supersymmetric gauged
WZW models. In the long run, it should perhaps lead to new insights
in the $N=2$ case since it is well-known that some points of the modular space
of the Calabi-Yau compactifications
 can be indeed written as (an orbifold
of) the tensor product of the (quantum) supersymmetric gauged WZW models.
In the $N=2$ WZW model the Poisson-Lie T-duality was shown to have the
same effect on the supercurrents as the usual mirror map \cite{P2}
and it was conjectured there that this should be the case also for
the Kazama-Suzuki models \cite{KS}.
It would be eventually interesting to identify the
 Poisson-Lie T-duality
with the mirror map also in the Calabi-Yau case.

In what follows, we shall shortly review the theory of the dressing cosets
 \cite{KS4}. Then we shall present the particular construction
which gives the gauged WZW model for a large class of groups $G$ and their
subgroups $H$. The dressing coset duality in this context will "intersect"
with the Kiritsis-Obers duality. Finally, we shall give the supersymmetric
generalization of the bosonic picture.

 2. Consider  a $2n$-dimensional Lie group $D$
whose Lie algebra $\D$ is
equipped with a symmetric invariant non-degenerate
 bilinear form $\ll.,.\gg$.
It is required that $D$ also possess
 two $n$-dimensional non-conjugated subgroups $G,\ti G$ such that their
 Lie algebras
$\G,\ti G$
are isotropic.
Consider now an $n$-dimensional linear subspace $\E\ss\D$ such
that
it intersects with  its orthogonal
complement $\E^\perp$ in an isotropic Lie algebra $\F$, i.e.
\be \E\cap\E^\perp=\F; \qquad [\F,\F]\ss\F.\ee
Moreover, both $\E$ and $\E^\perp$ should be invariant subspaces with respect
to the adjoint action of $\F$:
\be [\F,\E]\ss\E,\qquad  [\F,\E^\perp]\ss\E^\perp.\ee
It was shown in \cite{KS4} that all these data define a
 pair of dual non-linear
$\si$-models living respectively, on the targets $F\backslash D/G$ and
$F\backslash D/\ti G$. Their common dynamics is encoded in the first order
Hamiltonian action \cite{KS4}
$$ S=\jp\int {\rm d}\si{\rm d}\tau \ll \ds ll^{-1},\dt ll^{-1}\gg+
{1\over 12}\int d^{-1}\ll dll^{-1}[dll^{-1},dll^{-1}]\gg $$
\be -
\jp\int {\rm d}\si{\rm d}\tau \{\ll (\ds ll^{-1})_0,(\ds ll^{-1})_0\gg
-\ll (\ds ll^{-1})_1,(\ds ll^{-1})_1\gg\}.\ee
Here $l(\si,\tau)=l(\si+2\pi,\tau)$ is a mapping from a cylindrical worldsheet
into the group manifold $D$  and $\ds ll^{-1}$
is constrained to lie in $\F^\perp$:
\be \ds ll^{-1}\in\F^\perp.\ee
Note that due to the non-degeneracy of the form $\ll .,.\gg$ we have
\be\F^\perp={\rm Span}(\E,\E^\perp).\ee
We should also explain the meaning of the subscripts $0$ and $1$ in (3).
We can write arbitrary element
 $x\in\F^\perp$ as
\be x=x_0+x_1, \qquad x_0\in\E,\quad x_1\in\E^\perp.\ee
Of course, this decomposition is not unique, because the linear spaces
$\E$ and $\E^\perp$ intersect at $\F$. The decomposition
$ x=x_0' +x_1'$, where $x_0'=x_0+\phi$ and
$x_1'=x_1-\phi$,  is equally good for an arbitrary $\phi\in\F$. However,
due to the fact that $\ll\F,\E\gg=\ll\F,\E^\perp\gg=0$, the action (3)
does not depend on this decomposition.

The action (3) possesses the following bosonic gauge symmetry
$l\to fl, f(\si,\tau)\in F$ which explains why we take the left coset
 $F\backslash D/G$.

The way how to obtain the dual pair of the $\si$-models from the action (3)
was described in \cite{KS2}. Here we shall review it for the particular
choice of the Drinfeld double and its subgroups which leads to the gauged
WZW model.
The double is the direct product $D= G\times G$
where $G$ is a real Lie group whose Lie algebra
$\G$  is itself equipped with a symmetric
invariant nondegenerate bilinear form.  Note, however, that in distinction
with the group $D$, the form $<.,.>$ need not have any null vectors. This
is the case, for instance,  for simple compact groups and their
 Killing-Cartan forms.  We shall moreover require an existence of an outer
automorphism $\chi$ of $G$ such that it preserves the bilinear form.
For example, all groups $SU(n); n\geq 3$ fulfil our requirements.
The bilinear form $\ll.,.\gg$ on the Lie algebra $\D$ of $D$ for any pair
of elements $(X,Y),(U,V)\in \D$ is given by
\be
\ll(X,Y),(U,V)\gg=<X,U>-<Y,V>.
\ee
Two maximally isotropic subgroups $G^\delta\ss\D$ and
 $G^\chi\ss\D$ are defined as follows
\be G^\delta =\{(g,g)\in\D; g\in G\},
\quad G^\chi=\{(\chi(g),g)\in\D; g\in G\}.\ee

Now let us fix some Lie subgroup $H$ of $G$ such that the restriction of
 $<.,.>$
on its Lie algebra $\H$ is non-degenerate.
The subspaces $\E, \E^{\perp}$ defining the dressing coset models
on $F\backslash D/G^\delta$ and on $F\backslash D/G^\chi$ are given
as follows

\be
\E=(\H^{\perp},0)\oplus (\H,\H);\quad
\E^{\perp}=(0,\H^{\perp})\oplus (\H,\H).
\ee
The subalgebra $\F$ and the subspace  $\F^{\perp}$ are given by
\be
\F=(\H,\H),\quad
\F^{\perp}=(\H^{\perp},0)\oplus (\H,\H)\oplus (0,\H^\perp).
\ee
We note immediately that $F=H^\delta\ss\D$ and that the conditions (1)
and (2) are satisfied.  Hence we have a special case of the dressing
coset construction.

It is easy to identify the  target of the model
 $F\backslash D/G^\delta =
H^\delta\backslash G\times G/
G^\delta$. First of all, we note that every element $l$ of $G\times G$
can be written as
\be l=(k,1)(g,g); \quad k,g\in G.\ee
Hence the simple coset $G\times G/G^\delta$ can be obviously identified
with $G$. Let us now act from the left on $l$ by some element
$(h,h)\in H^\delta$. We obtain
\be (h,h)(k,1)(g,g)=(hkg,hg)=(hkh^{-1},1)(hg,hg).\ee
From here we conclude that the target is simply $G/Ad(H)$, which is the
same thing as the target of the gauged $G/H$ WZW model. In the same
way we can find the dual target $H^\delta\backslash G\times G/
G^\chi$. It is given by the cosets of the following
action of $H$ on $G$:
$^h g=hg\chi(h)^{-1}$. If $H$ is stable with respect to the action
of the automorphism $\chi$ and the restriction of $\chi$ on $H$
is an outer automorphism of $H$, we obtain the target of the "axially" gauged
 WZW model  in the sense of Bars \& Sfetsos and Kiritsis \& Obers \cite{BS,KO}.
 If $H$ is not stable
 then we are in the framework of Chung \& Tye \cite{CT}.

In the setting (9-10)  and with the ansatz (11)
 the action (3) of our dressing coset becomes
$$S[k,g]=\jp\int {\rm d}\tau {\rm d}\si <\ds kk^{-1},\dt k k^{-1}>
+{1\over 12}\int d^{-1}<dkk^{-1},[dkk^{-1},dkk^{-1}]>$$
$$+\int {\rm d}\tau {\rm d}\sigma\{<k^{-1}\partial_{\tau}k,
\partial_{\sigma}gg^{-1}>
-\jp<k^{-1}\partial_{\sigma}k,k^{-1}\partial_{\sigma}k>$$
\be -<k^{-1}\partial_{\sigma}k,\partial_{\sigma}gg^{-1}>
 -<P^{\perp}\partial_{\sigma}gg^{-1},\partial_{\sigma}gg^{-1}>\}.\ee
In deriving (13), we have used the Polyakov-Wiegmann formula \cite{PW}
and the fact that $\G^\delta$ is isotropic.
The notations :
$P, P^{\perp}$ are the projectors on the subspaces $H$ and $H^{\perp}$
correspondingly; $Ad(k)X=kXk^{-1}$.

The constraint (4) becomes:
\be
P\partial_{\sigma}gg^{-1}=
(1-PAd(k)P)^{-1}P(\partial_{\sigma}kk^{-1}
+Ad(k)P^{\perp}\partial_{\sigma}gg^{-1}) .
\ee

Inserting $P\ds gg^{-1}$ from (14) back into the action we obtain
an expression which depends on $g$ only via $P^\perp \ds gg^{-1}$:
$$S[k,g]=\jp\int {\rm d}\tau {\rm d}\si <\ds kk^{-1},\dt k k^{-1}>
+{1\over 12}\int d^{-1}<dkk^{-1},[dkk^{-1},dkk^{-1}]>$$
$$+\int {\rm d}\tau {\rm d}\sigma
\{<k^{-1}\partial_{-}k,
P^{\perp}\partial_{\sigma}gg^{-1}$$
$$+\jp (1-PAd(k)P)^{-1}P(\partial_{\sigma}kk^{-1}
+Ad(k)P^{\perp}\partial_{\sigma}gg^{-1})>$$\be
-\jp<k^{-1}\partial_{\sigma}k,k^{-1}\partial_{\sigma}k>
 -<P^{\perp}\partial_{\sigma}gg^{-1},\partial_{\sigma}gg^{-1}>\}.
\ee
After solving away  the field $P^{\perp}\partial_{\sigma}gg^{-1}$,
we finally obtain the action:
$$
S[k]=\int {\rm d}\xi^+ {\rm d}\xi^-
{1\over 2}<\partial_{-}kk^{-1},\partial_{+}kk^{-1}>+
{1\over 12}\int d^{-1}<dkk^{-1},[dkk^{-1},dkk^{-1}]>$$
\be +\int {\rm d}\xi^+ {\rm d}\xi^-
<k^{-1}\partial_{-}k,
(1-PAd(k)P)^{-1}P\partial_{+}kk^{-1}>.
\ee
This is the same action that  we  obtain if we integrate out the gauge
fields $A_{+},A_{-}\in\H$ from the gauged WZW action of the coset model
$G/Ad(H)$:

$$ S[k]=\jp\int {\rm d}\xi^{+}{\rm d}\xi^{-}<\partial_+ kk^{-1},
\partial_- kk^{-1}>+
{1\over 12}\int d^{-1}<dkk^{-1},[dkk^{-1},dkk^{-1}]>$$
$$ -2\int {\rm d}\xi^{+}{\rm d}\xi^{-}
\{<\partial_{+}kk^{-1},A_{-}>$$\be -<k^{-1}\partial_{-}k,A_{+}>
-<A_{-},A_{+}>+<A_{-},Ad(k)A_{+}>\},
\ee
note that we have used here the light-cone variables of integration
\be
{\rm d}\xi^{\pm}=({\rm d}\tau\pm {\rm d}\sigma)/2.\ee

The action of the model dual to (16) can be obtained from the duality
invariant action (3) by using the ansatz
\be l=(k,1)(\chi(g),g)\ee
and solving away the field $g$. The result is the action
$$
S[k]=\jp\int {\rm d}\xi^+ {\rm d}\xi^-
<\partial_{-}kk^{-1},\partial_{+}kk^{-1}>+
{1\over 12}\int d^{-1}<dkk^{-1},[dkk^{-1},dkk^{-1}]>$$
\be +\int {\rm d}\xi^+ {\rm d}\xi^-
\{<(\chi^{-1})^* k^{-1}\partial_{-}k,
(1-PAd(k)\chi^* P)^{-1}P\partial_{+}kk^{-1}>\},
\ee
where $\chi^*$ denotes the push forward map $\G\to\G$.

3. The supersymmetric case. Strictly speaking, the theory of the
supersymmetric dressing cosets has not been so far discussed in the literature.
On the other hand, it is not difficult to develop it since the principal
tool, the duality invariant action for the
supersymmetric $D/G\leftrightarrow D/\ti G$ duality, has been found already in
\cite{K}. Thus we find after some work that the action for the supersymmetric
 dressing cosets is  given by
$$
S[l,\psi^{\pm}]={1\over 2}
\int {\rm d}\si{\rm d}\tau \ll\partial_{\sigma}ll^{-1},
\partial_{\tau}ll^{-1}\gg
+{1\over 12}\int d^{-1}\ll dll^{-1},[dll^{-1},dll^{-1}]\gg $$
$$
\int {\rm d}\si{\rm d}\tau \{-\js\ll \psi^{+},\partial_{-}\psi^{+}\gg
+\js\ll \psi^{-},\partial_{+}\psi^{-}\gg
+{1\over 8}\ll \psi^{+}\psi^{+} ,\psi^{+}\psi^{+}\gg$$
$$-{1\over 8}\ll \psi^{-}\psi^{-} ,\psi^{-}\psi^{-}\gg
+\jp\ll\partial_{\sigma}ll^{-1},\psi^{+}\psi^{+}+\psi^{-}\psi^{-}\gg $$
$$-{1\over 8}\ll (\{\psi^{-},\psi^{+}\})_0,(\{\psi^{-},\psi^{+}\})_0\gg
+{1\over 8}\ll (\{\psi^{-},\psi^{+}\})_1,(\{\psi^{-},\psi^{+}\})_1\gg
$$  $$-\jp\ll(\partial_{\sigma}ll^{-1}+{1\over 2}\psi^{+}\psi^{+}
-{1\over 2}\psi^{-}\psi^{-})_0,(\partial_{\sigma}ll^{-1}+
{1\over 2}\psi^{+}\psi^{+}
- {1\over 2}\psi^{-}\psi^{-})_0\gg $$
\be +\jp\ll(\partial_{\sigma}ll^{-1}+{1\over 2}\psi^{+}\psi^{+}
-{1\over 2}\psi^{-}\psi^{-})_1,(\partial_{\sigma}ll^{-1}+
{1\over 2}\psi^{+}\psi^{+}
- {1\over 2}\psi^{-}\psi^{-})_1\gg \}.\ee
The analogue of the constraint (4) is now
\be  \partial_{\sigma}ll^{-1}+{1\over 2}\psi^{+}\psi^{+}
-{1\over 2}\psi^{-}\psi^{-}\in\F^\perp .\ee
Moreover,
 the fermions $\psi^\pm$ fulfil
\be \psi^+\in\E,\quad \psi^-\in\E^\perp\ee
and the consistency requires that
\be [\E,\E^\perp]\ss\F^\perp.\ee
The latter requirement is indeed fulfilled for our particular choice  (9-10).

The action (21) possesses the following bosonic gauge symmetry
\be l\to fl, \quad \psi^\pm\to f\psi^\pm f^{-1},\quad f(\si,\tau)\in F;\ee
and the following fermionic gauge symmetry
\be \psi^\pm \to \psi^\pm+\xi^\pm, \quad \xi^\pm (\si,\tau)\in\F.\ee
The target spaces of the corresponding dual pair of the $\si$-model
are the same as in the bosonic case.

Now for our concrete case (9) and (10), we can parametrize the fermions as
\be \psi^{+}=(\gamma^{+},0)+(\eta^{+},\eta^{+}),\quad
\psi^{-}=(0,\gamma^{-})+(\eta^{-},\eta^{-}),\ee
where $\gamma^\pm\in\H^\perp$ and $\eta^\pm\in\H$.

The fermionic gauge symmetry (25) can be used to set
$\eta^\pm =0$. In what follows, we shall always use this particular gauge.
Then the action of our supersymmetric dressing coset is
$$
S[l,\psi^{\pm}]=\jp\int d\tau d\sigma
<\partial_{\sigma}kk^{-1},\partial_{\tau}kk^{-1}> +{1\over 12}\int d^{-1}<
dkk^{-1},[dkk^{-1},dkk^{-1}]>)$$
$$+\int d\tau d\sigma\{<k^{-1}\partial_{\tau}k,\partial_{\sigma}gg^{-1}>
-{1\over 4}(<\gamma^{+},\partial_{-}\gamma^{+}>
+<\gamma^{-},\partial_{+}\gamma^{-}>)$$
$$-<\partial_{\sigma}gg^{-1}
-{1\over 4}\gamma^{-}\gamma^{-},P\gamma^{-}\gamma^{-}>
-\jp <k^{-1}\partial_{\sigma}k,k^{-1}\partial_{\sigma}k>$$
\be -<k^{-1}\partial_{\sigma}k,\partial_{\sigma}gg^{-1}>
-<P^{\perp}\partial_{\sigma}gg^{-1},\partial_{\sigma}gg^{-1}>\}\ee
with the constraint
\be P(\partial_{\sigma}kk^{-1}-(1-Ad(k))\partial_{\sigma}gg^{-1}
+{1\over 2}(\gamma^{+}\gamma^{+}+\gamma^{-}\gamma^{-}))=0.
\ee
Now one first expresses $P\ds gg^{-1}$ from the constraint (28) and inserts
it back into the action (27). Then one can solve away also
the field $P^{\perp}\partial_{\sigma}gg^{-1}$.
As a result we obtain the action of the nonabelian coset model $G/Ad(H)$.
$$  S[k,\gamma^{\pm}]=\jp\int {\rm d}\xi^+ {\rm d}\xi^-
<\partial_{-}kk^{-1},\partial_{+}kk^{-1}>+
{1\over 12}\int d^{-1}<dkk^{-1},[dkk^{-1},dkk^{-1}]>$$
$$+\int {\rm d}\xi^+ {\rm d}\xi^-
\{-{1\over 2}(<\gamma^{+},\partial_{-}\gamma^{+}>
+<\gamma^{-},\partial_{+}\gamma^{-}>)$$
\be +<k^{-1}\partial_{-}k-\gamma^{-}\gamma^{-},
(1-PAd(k)P)^{-1}P(\partial_{+}kk^{-1}+\gamma^{+}\gamma^{+})>\}.
\ee
This is the same action we would obtain if to integrate out the gauge
fields $A_{+},A_{-}$ from the gauged WZW action of the coset model $G/Ad(H)$:

$$ S[k,\gamma^{\pm}]=\jp\int {\rm d}\xi^+ {\rm d}\xi^-
<\partial_{-}kk^{-1},\partial_{+}kk^{-1}>+
{1\over 12}\int d^{-1}<dkk^{-1},[dkk^{-1},dkk^{-1}]$$
$$-2\int {\rm d}\xi^+ {\rm d}\xi^-
\{{1\over 2}(<\gamma^{+},\partial_{-}\gamma^{+}+[A_{-},\gamma^{+}]>$$
$$+<\gamma^{-},\partial_{+}\gamma^{-}+[A_{+},\gamma^{-}] >)
 +<\partial_{+}kk^{-1},A_{-}>$$\be -<k^{-1}\partial_{-}k,A_{+}>
-<A_{-},A_{+}>+<A_{-},Ad(k)A_{+}>\}.
\ee
The action of the model dual to (29) reads
$$  S[k,\gamma^{\pm}]=\jp\int  {\rm d}\xi^+ {\rm d}\xi^-
<\partial_{-}kk^{-1},\partial_{+}kk^{-1}>+
{1\over 12}\int d^{-1}<dkk^{-1},[dkk^{-1},dkk^{-1}]>$$
$$+\int {\rm d}\xi^+ {\rm d}\xi^-
\{-{1\over 2}(<\gamma^{+},\partial_{-}\gamma^{+}>
+<\gamma^{-},\partial_{+}\gamma^{-}>)$$
\be +<(\chi^{-1})^* k^{-1}\partial_{-}k-\gamma^{-}\gamma^{-},
(1-PAd(k)\chi^*P)^{-1}P(\partial_{+}kk^{-1}+\gamma^{+}\gamma^{+})>\}.
\ee

\end{document}